\documentclass[aip,reprint,amsmath,amssymb]{revtex4-1}

\usepackage{graphicx}
\usepackage{dcolumn}
\usepackage{bm}

\usepackage[utf8]{inputenc}
\usepackage[T1]{fontenc}
\usepackage{mathptmx}

\usepackage{hhline}
\usepackage[normalem]{ulem}

\usepackage{color}

\begin{document}

\title{Hybrid optical fiber for light-induced superconductivity}

\author{Evgeny Sedov}
\email[Electronic address: ]{evgeny\_sedov@mail.ru}
\affiliation{Westlake University, 18 Shilongshan Road, Hangzhou 310024, Zhejiang Province, China}
\affiliation{Institute of Natural Sciences, Westlake Institute for Advanced Study, 18 Shilongshan Road, Hangzhou 310024, Zhejiang Province, China}
\affiliation{Department of Physics and Applied Mathematics, Vladimir State University named after A. G. and N. G. Stoletovs, Gorky str. 87, 600000, Vladimir, Russia}
\author{Irina Sedova}
\affiliation{Department of Physics and Applied Mathematics, Vladimir State University named after A. G. and N. G. Stoletovs, Gorky str. 87, 600000, Vladimir, Russia}
\author{Sergey Arakelian}
\affiliation{Department of Physics and Applied Mathematics, Vladimir State University named after A. G. and N. G. Stoletovs, Gorky str. 87, 600000, Vladimir, Russia}
\author{Giuseppe Eramo}
\affiliation{Mediterranean Institute of Fundamental Physics, 31, Appia Nuova, Frattocchi, Rome, 00031, Italy}
\author{Alexey Kavokin}
\affiliation{Westlake University, 18 Shilongshan Road, Hangzhou 310024, Zhejiang Province, China}
\affiliation{Institute of Natural Sciences, Westlake Institute for Advanced Study, 18 Shilongshan Road, Hangzhou 310024, Zhejiang Province, China}
\affiliation{Spin Optics Laboratory, St. Petersburg State University, Ul’anovskaya 1, Peterhof, St. Petersburg 198504, Russia}

\begin{abstract}
We exploit the recent proposals for the light-induced superconductivity mediated by a Bose-Einstein condensate of exciton-polaritons to design a superconducting fiber that would enable long-distance transport of a supercurrent at elevated temperatures.
The proposed fiber consists of a conventional core made of a silica glass with the first cladding layer formed by a material sustaining dipole-polarised excitons with a binding energy exceeding 25 meV.
To be specific, we consider a perovskite cladding layer of 20 nm width.
The second cladding layer is made of a conventional superconductor such as aluminium.
The fiber is covered by a conventional coating buffer and by a plastic outer jacket.
We argue that the critical temperature for a superconducting phase transition in the second cladding layer may be strongly enhanced due to the coupling of the superconductor to a bosonic condensate of exciton-polaritons optically induced by the evanescent part of the guiding mode confined in the core.
The guided light mode would penetrate to the first cladding layer and provide the strong exciton-photon coupling regime.
We run simulations that confirm the validity of the proposed concept.
The fabrication of superconducting fibers where a high-temperature superconductivity could be controlled by light would enable passing superconducting currents over extremely long distances.
\end{abstract}


\maketitle


The theoretical concept of conventional superconductivity introduced by Bardeen, Cooper and Schrieffer (BCS)~\cite{PhysRev106162} relies on the pairing of electrons in a Fermi sea due to the exchange by quanta of crystal lattice vibration: acoustic phonons.
Since 1970s, multiple attempts were made to replace phonons by a more efficient binding agent that would strengthen electron-electron attraction and enable superconductivity at higher temperatures.
Excitons have been put forward by Allender, Bray and Bardeen~\cite{PhysRevB71020} and Ginzburg~\cite{UFN1183151976} as promising candidates for playing the role of such a binding agent in hybrid metal-semiconductor structures.
However, despite of significant efforts to fabricate structures where the exciton-mediated superconductivity would be observable, no experimental evidence for this mechanism was reported till now, to the best of our knowledge.
The interest to exciton-mediated superconductivity was renewed after the discovery of the Bose-Einstein condensation of light-matter bosonic quasiparticles, exciton-polaritons, in semiconductor microcavities, initially at the liquid Helium temperature~\cite{Nature4434092006} and then at the room temperature~\cite{PhysRevLett98126405}.
Exciton-polaritons formed by strongly coupled elementary crystal excitations (excitons) and cavity photons~\cite{Kavokin2017Book} accumulate by tens thousands in a single quantum state thus giving rise to the polariton-lasing~\cite{PhysRevLett98126405}.
In a series of theoretical works,~\cite{PhysRevLett104106402,JNP6064502,CHEROTCHENKO2016170,PhysRevB93054510} it was shown that the condensates of exciton polaritons may interact with free electrons much stronger than individual virtual excitons considered in the early works~\cite{PhysRevB71020,UFN1183151976}.
This paves the way to the realisation of exciton-mediated superconductivity in hybrid multilayer structures where a free electron gas would be placed in the vicinity of a bosonic condensate of exciton-polaritons.
The observation of Bose-Einstein condensation of exciton-polaritons at the room temperature~\cite{PhysRevLett98126405,Kavokin2017Book} encouraged efforts for the observation of superconductivity mediated by exciton-polaritons at elevated temperatures.
It has been argued~\cite{JNP6064502} that a stationary dipole polarisation of excitons in the condensates might help maximising the strength of exciton-electron coupling, which is why strongly coupled structures containing spatially indirect excitons might be suitable for the realisation of exciton-mediated superconductivity.
A variety of materials potentially promising from this point of view has been considered, including the two-dimensional monolayers of transition metal dichalcogenides~\cite{PhysRevB93054510,NatMat155992016} and perovskite nanoplatlets~\cite{NanoLett1739822917}.
In Ref.~\onlinecite{PhysRevLett120107001} it has been argued that the interplay between conventional BCS superconductivity and the exciton-mediated superconductivity may result in the resonant enhancement of the critical temperature for the superconducting phase transition $T_{\text{c}}$, thus, potentially, paving the way to room temperature superconductivity that would be fully controlled by light that is used to pump exciton-polariton condensates.
The recent experimental studies~\cite{Science3311892011,Nature5304612016} revealed features of the light-induced superconductivity.
It was argued that the phase transition has been triggered by an optically pumped vibron mode that represents a similar mechanism to the exciton-induced superconductivity.
Still, no stationary increase of $T_{\text{c}}$ mediated by laser illumination has been reported so far, to our knowledge.

\begin{figure}[!t]
\begin{center}
\includegraphics[width=.99\linewidth]{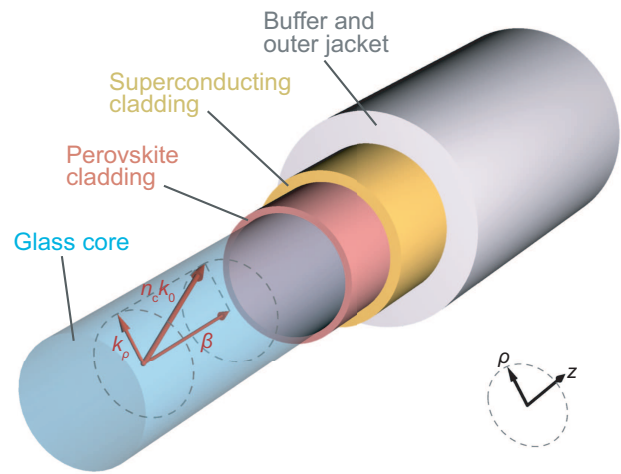}
\end{center}
\caption{\label{FIG_Scheme}
The schematic of a proposed design.
The bosonic condensate of dipole polarised exciton-polaritons is formed by the optical pumping through the guided optical modes of the fiber.
The strong exciton-photon coupling regime is achieved due to the overlap of the photon mode localised in the core with the exciton state located in the first cladding layer (perovskite).
The proximity of the perovskite layer to the second cladding layer (superconductor) ensures the efficient coupling of the dipole-polarised condensate of polaritons with the electron Fermi sea in the superconductor.
This coupling leads to a significant increase of the critical temperature of the superconducting phase transition.
The core and cladding layers are protected by plastic buffer and jacket. 
The red arrows indicate the decomposition of a guided mode wave vector on the transversal component $\mathbf{k}_{\rho}$ and the propagation constant $\beta$ along the main axis of the fiber.
}
\end{figure}

In this Letter, we propose a design of the structure that would enable light-mediated superconductivity triggered by a bosonic condensate of exciton-polaritons in a conventional optical fiber~\cite{TONG20124641} containing two cladding layers:
one made of a material sustaining dipole-polarised excitons that are able to strongly couple to the guided optical mode, and the second one made of a conventional superconductor (see the schematic in Fig.~\ref{FIG_Scheme}).
To be specific, we consider a first cladding layer made of a perovskite material where polariton lasing was recently demonstrated in micro- and nanowires~\cite{adom201701032,NatMater146362015}, and a conventional superconductor aluminium.~\cite{JETP431173}
We note that the exciton binding energy in the considered perovskite material exceeds 25~meV which makes it suitable for the room temperature operation.
The critical temperature for the superconducting phase transition in aluminium is about 2~K in the dark, but it may be strongly enhanced by the coupling to the polariton condensate.
From the theoretical point of view, the novelty of the considered design is in the replacement of a stationary polariton condensate considered in the previous studies~\cite{PhysRevLett104106402,JNP6064502,CHEROTCHENKO2016170,PhysRevB93054510} with a moving condensate that may be treated as a coherent quantum liquid~\cite{RevModPhys85299}.
From the practical point of view, the realisation of optical fibers with superconducting cladding layers would pave the way to the long-range transport of supercurrents.
The coupling of a superconductor with a condensate of exciton-polaritons that is pumped by a laser light passing through the core would constitute a tool for the enhancement of $T_{\text{c}}$ as well as the switching mechanism that enables the optical control of superconductivity.
It is important to note that while the absorption of light in a superconducting cladder is inevitable, in principle, it can be minimized by proper designing the fiber and compensated due to the pumping of polaritons through one of the higher frequency guided light modes of the fiber.
Below we present estimations of the characteristics of the superconducting hybrid fiber and provide specific recommendations for its design.

The proposed hybrid optical fiber is schematically shown in Fig.~\ref{FIG_Scheme}.
The cylindrical core is assumed being made of a conventional silica glass.
For further estimations, we take the refractive index of the core as $n_{\text{C}} = 1.45$.
The first cladding layer intended to be a holder of excitons is a perovskite layer.
Among a variety  of perovskites, we give a preference to methylammonium lead tribromide MAPbBr$_3$ (MA = CH$_3$NH$_3$).
The strong coupling of excitons with light in the waveguide geometry has been recently reported for the structures where MAPbBr$_3$ was used as a core nanowire~\cite{adom201701032,NatMater146362015}. 
The peculiar optical and dielectric properties of MAPbBr$_3$ in the spectral range from visible to near-ultraviolet have been extensively studied both experimentally and theoretically for last several years\cite{C5NR05435D,NatMater146362015,SmallMet117001632017} in the context of developing solar cell circuits and photonic devices.
In the photon energy range following immediately above the band gap starting around 2.24~eV and extending to about 4.5~eV, the MAPbBr$_3$ film is characterized by three absorption peaks at energies of about 2.3, 3.5 and 4.5~eV.~\cite{C5NR05435D,JAP1211155012017}
Authors of Ref.~\onlinecite{C5NR05435D} show that this peculiarity of the observed spectrum can be described by a dielectric function characterized by four resonances within the indicated energy region.
The lowest energy resonance situated near the band gap (at about $2.3$~eV) is split from the next one (situated at about 3.5~eV) by more than 1~eV, which significantly exceeds the characteristic interaction energies in our system. 
This allows one to use the coupled oscillator model of the perovskite dielectric function in the simulations:~\cite{adom201701032}
\begin{equation}
\varepsilon_{\text{P}}(\omega) = \varepsilon_{\text{b}} \left(1 + \frac{\omega_{\text{l}}^2 - \omega_{\text{t}}^2 }{\omega_{\text{t}}^2 - \omega^2 - \text{i} \omega \Gamma} \right),
\end{equation}
where $ \varepsilon_{\text{b}}$ is the background dielectric constant, $\omega_{\text{l},\text{t}}$ are the longitudinal and transverse exciton frequencies, $\Gamma$ is the non-radiative exciton damping.
Values of the parameters for the best fit of $\varepsilon_{\text{P}}(\omega)$ are following:~\cite{adom201701032} $\varepsilon_{\text{b}} = 4.7$, $\hbar \omega_{\text{l}} = 2.328$~eV, $\hbar \omega_{\text{t}} = 2.303$~eV and $\hbar \Gamma = 59$~meV. 
We take the thickness of the MAPbBr$_3$ cladding layer as 20~nm. 

When choosing the second (superconducting) cladding layer, we follow the original paper on the polariton-mediated superconductivity~\cite{PhysRevLett120107001} and consider for this role a thin film of aluminium.~\cite{JETP431173}
Aluminium possesses an advantage of high reflectivity properties, which contributes to reducing losses of the optical modes and strong coupling of the latter to excitons in the perovskite cladding layer.
In the simulations, we consider an aluminium film of thickness of 30~nm with the spectral dependences of the refractive index and the extinction coefficient taken from Ref.~\onlinecite{ACSPhot23262015}
The cladding layers can be separated from one another by a spacer of widths of the order of 1~nm with a refractive index matching that of the core.

To reach the effective coupling of the perovskite exciton with a light mode in the waveguide,  we should bring one of the optical guided modes into the resonance with the exciton energy.
This imposes restrictions on the wave number $\beta$ characterizing the propagation of the guided mode along the main axis of the waveguide, see Fig.~\ref{FIG_Scheme}.
For the bare waveguide representing a glass core in air, the condition for $\beta$ in general form is given by $k_0 < \beta < n_{\text{C}}k_0$ with $k_0 = \omega_0/c$ and $\omega _0$ being the wave number and the frequency of light in air.
A guided mode characterized by a frequency of $\omega = \omega_{\text{l}}$ should possess the wave number $\beta$ belonging to the range of $(11.7 \, \mu \text{m} ^{-1}, 16.9  \, \mu \text{m} ^{-1} )$.
Our simulations show that additional cladding layers extend this range by at least $3.3\, \mu \text{m} ^{-1}$.

For our hybrid optical fiber, we propose to take a glass core of a reasonably small diameter being of the order of the guided mode wavelength, $d_{\text{C}} \gtrsim \lambda$.
In such a waveguide, the higher energy guided modes are split from the ground mode by hundreds of millielectron volts.
It allows one to neglect the effect of the higher modes on the coupling of the fundamental mode to the perovskite excitons.
We note that higher energy guided modes are crucial for the long distance transmission of polariton superfluids and light-induced superconducting currents.
They serve to feed the polariton condensate and compensate for inevitable losses.
The amplification of the polariton mode may be done using the schemes used in optical fiber amplifiers based e.g. on the electronic transitions in Er atoms.~\cite{PayneFiber1990,IEEE413421992,IEEE413451992}

\begin{figure}[!t]
\begin{center}
\includegraphics[width=.96\linewidth]{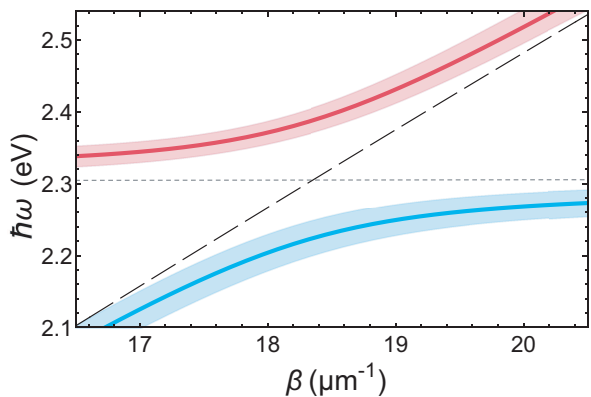}
\end{center}
\caption{\label{FIG_Disp}
The variation of the energy of exciton-polariton modes in the hybrid optical fiber with the increase of the propagation constant $\beta$.
The pale color shadows framing the curves indicate the decay rates of the modes.
The inclined dashed line shows the energy of the guided mode without the exciton resonance.
The horizontal dashed line indicates the perovskite exciton energy.
The core diameter is taken as $d_{\text{C}} = 0.8\, \mu\text{m}$.
}
\end{figure}

To examine modes of the hybrid optical fiber, we use the well-known transfer matrix method~\cite{Chew1995Book,Kavokin2017Book,PhysRevB94125309}, which is based on matching the components of the electromagnetic field at the interfaces of the homogeneous layers using the Maxwell boundary conditions.
For the details of the method adapted for a cylindrical geometry, we refer to Refs.~\onlinecite{OptSpectr888712000,PhysRevB86235425,PhysRevB6113791}.

\begin{figure}[!t]
\begin{center}
\includegraphics[width=.97\linewidth]{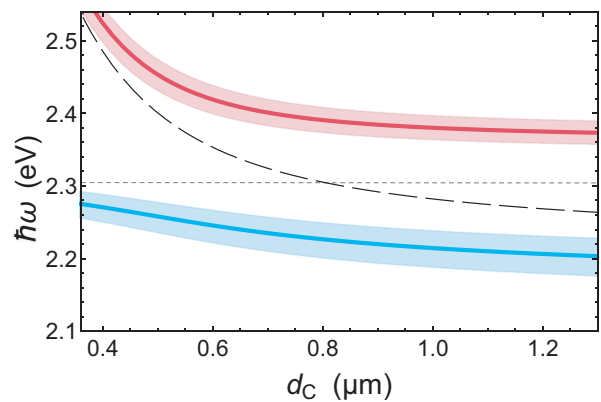}
\end{center}
\caption{\label{FIG_EonR}
The dependence of the energy of exciton-polariton modes in the hybrid optical fiber on its core diameter $d_{\text{C}}$.
The pale color shadows framing the curves indicate the decay rates of the modes.
The dashed curve shows the energy of the guided mode in the absence of the exciton resonance.
The horizontal dashed line indicates the perovskite exciton energy.
The propagation constant is taken as $\beta = 18.4\, \mu\text{m}^{-1}$.
}
\end{figure}

The dependence of the energy of the fundamental mode on the propagation constant $\beta$ in the hybrid optical fiber with the core diameter of $0.8\, \mu \text{m}$ is shown in Fig.~\ref{FIG_Disp}.
In the vicinity of the exciton resonance at $ \omega \approx \omega _{\text{t}}$ and $\beta \approx 18.4 \, \mu \text{m}^{-1}$ a clear Rabi splitting by $2 \hbar \Omega _{\text{R}} \approx $ 165~meV of the dispersion curve $\omega (\beta)$ into two branches is apparent, which results from the anti-crossing of the exciton and the guided optical mode dispersions.
The dispersions of the latter being linear in $\beta$ are shown in Fig.~\ref{FIG_Disp} by the dashed lines.
The anti-crossing of the dispersions is the manifestation of the appearance of the coupled exciton-photon states, exciton polaritons, which we consider for the role of mediators of the superconductivity.
An indispensable condition for the appearance of polaritons is the strong coupling condition, which implies that the characteristic losses in the system should not exceed the splitting $2 \hbar \Omega _{\text{R}}$.
The color shadows framing the dispersion curves in Fig.~\ref{FIG_Disp} show the broadening of the polariton modes.
One can see that the Rabi spitting $2 \hbar \Omega _{\text{R}}$ exceeds the losses with a large margin.

Figure~\ref{FIG_EonR} shows the dependence of the exciton-polariton energy on the diameter of the core $d_{\text{C}}$ for the polariton modes characterized by the wave number $\beta = 18.4\, \mu\text{m}^{-1}$.
The energy of polaritons of both branches increases with the decreasing diameter $d_{\text{C}}$.
The predominance of the Rabi splitting $2 \hbar \Omega _{\text{R}}$ over the characteristic losses on the entire considered range of $d_{\text{C}}$ keeps the strong coupling condition fulfilled. 
By choosing the diameter of the glass core, one can tune characteristics of the guided polariton mode in the $\omega \, - \, \beta$ plane.

In summary, we have proposed the structure of a hybrid optical fiber intended to provide an effective interaction of a superconductor with exciton polaritons, guided modes modified by coupling with excitons in a perovskite cladding layer.
We have demonstrated that the strong-coupling regime for exciton polariton formation is realizable in this geometry. 
Exciton polaritons will fulfil the role of mediators of coupling of electrons in a superconductor which is expected to facilitate elevating the critical temperature of superconductivity.

\section*{Acknowledgements}

This work is from the Innovative Team of International Center for Polaritonics and is supported by Westlake University (Project No. 041020100118).
E.S. and I.S. acknowledge partial support from the Grant of the President of the Russian Federation for state support of young Russian scientists  No. MK-2839.2019.2.
A.K. acknowledges the Saint-Petersburg State University for the research grant ID 40847559.

\bibliography{waveguideBibl}

\end{document}